\documentclass{aastex61}

\usepackage{mathrsfs}
\DeclareMathAlphabet{\mathpzc}{OT1}{pzc}{m}{it}

\newcommand{\bea}{\begin{eqnarray}}
\newcommand{\eea}{\end{eqnarray}}

\received{April 11, 2017}
\revised{\today}
\accepted{\today}
\submitjournal{ApJ}

\shorttitle{Induced gravitational waves}
\shortauthors{Hwang et al.}

\begin{document}

\title{Gauge dependence of gravitational waves generated from scalar perturbations}

\correspondingauthor{Donghui Jeong}
\email{djeong@psu.edu}

\author{Jai-chan Hwang}
\affiliation{
Department of Astronomy and Atmospheric Sciences, Kyungpook National University, Taegu, Korea \\
}

\author[0000-0002-8434-979X]{Donghui Jeong}
\affiliation{Department of Astronomy and Astrophysics and Institute for Gravitation and the Cosmos,
The Pennsylvania State University, University Park, PA 16802, USA
}

\author{Hyerim Noh}
\affiliation{
Korea Astronomy and Space Science Institute, Daejon, Korea
}



\begin{abstract}

A tensor-type cosmological perturbation, defined as a transverse and traceless spatial fluctuation, is often interpreted as the gravitational waves. While decoupled from the scalar-type perturbations in linear order, the tensor perturbations can be sourced from the scalar-type in the nonlinear order. The tensor perturbations generated by the quadratic combination of linear scalar-type cosmological perturbation are widely studied in the literature, but all previous studies are based on zero-shear gauge without proper justification. Here, we show that, being second order in perturbation, such an induced tensor perturbation is generically gauge dependent. In particular, the gravitational wave power spectrum depends on the hypersurface (temporal gauge) condition taken for the linear scalar perturbation. We further show that, during the matter-dominated era, the induced tensor modes dominate over the linearly evolved primordial gravitational waves amplitude for $k\gtrsim10^{-2}~[h/{\rm Mpc}]$ even for the gauge that gives lowest induced tensor modes with the optimistic choice of primordial gravitational waves ($r=0.1$). The induced tensor modes, therefore, must be modeled correctly specific to the observational strategy for the measurement of primordial gravitational waves from large-scale structure via, for example, parity-odd mode of weak gravitational lensing, or clustering fossils.

\end{abstract}



\section{Introduction}
                                         \label{sec:Introduction}

From the beginning of cosmological perturbation theory \citep{Lifshitz-1946},
it is well known that linear order relativistic perturbations can be decoupled
into the scalar-, vector- and tensor-type perturbations and the tensor-type
perturbations are gauge-invariant in the (spatially homogeneous and isotropic)
Friedmann background world model. The gauge dependence, especially the
temporal gauge (hypersurface or slicing) dependence, of the scalar-type
perturbations is also well known in the literature
\citep{Bardeen-1980}. It was Bradeen who suggested a
practical strategy of utilizing the gauge dependence as an advantage in
analyzing the perturbations. He wrote ``The moral is
that one should work in the gauge that is mathematically most convenient for
the problem at hand.'' \citep{Bardeen-1988} 

The natural question arises: among all possible gauge conditions, which one
is more relevant for interpreting the physical world?
The answer to this question does not depend on the mathematical
structure of the gauge: not on the gauge invariance of the variable in the
chosen gauge, nor on the explicit gauge-invariance of the combination of
perturbation variables. As a matter of fact, all perturbation variables without
the gauge-mode ambiguity are gauge-invariant in the senses that their values
evaluated in any other gauges remain the same. It is the case for all
perturbation variables in the several fundamental gauges introduced in
\citet{Bardeen-1980,Bardeen-1988} except for the synchronous gauge. For these
gauges, each perturbation variable uniquely corresponds to a gauge-invariant
combination of perturbation variables. Again, according to Bardeen ``While a useful tool, gauge-invariance in itself does not
remove all ambiguity in physical interpretation,'' and ``Many gauge-invariant
combinations of these scalars can be constructed, but for the most part they
have no physical meaning independent of a particular time gauge, or
hypersurface condition.'' \citep{Bardeen-1988} 
For a given variable, say density perturbation or velocity perturbation, we
can, in fact, construct the variable with infinitely many different gauge
conditions, and all of them correspond to the gauge-invariant combinations;
this is because the constant-time hypersurface can be deformed in a continuous
manner. These statements are also true to the nonlinear order in cosmological
perturbation theory \citep{Noh-Hwang-2004,Hwang-Noh-2013}. It is, therefore,
safe to treat a variable evaluated in different gauges as entirely different
variables.
Finally, according to Bardeen ``Gauge-invariant variables give mathematically unambiguous ways of comparing results obtained in different gauges, but their physical interpretation is not necessarily straightforward, in that it is usually tied to a particular way of slicing the spacetime into hypersurfaces. I know of no way to characterize completely the deviations from homogeneity and isotropy independent of the slicing into spacelike hypersurfaces.'' \citep{Bardeen-1988}

Instead, which gauge-invariant variable corresponds to the one that we measure
from observation depends on the nature of the observation. That is, a
specification of observation must tell us which gauge-invariant variable or
combinations of them is the right one for that particular observation
(assuming, of course, that perturbation theory in
the Friedmann world model handles the observed phenomena). We discuss this
issue further in Section \ref{sec:Discussion}.

To the nonlinear order, the three types of perturbations couple to each other
in the equation level and the decomposition itself becomes ambiguous.
It is because we can introduce many different ways of decomposing the
perturbation to scalar-, vector-, and tensor-types \citep{Hwang-Noh-2013};
hereafter we simply call it the scalar perturbation or scalar mode, etc.
Naturally, from the second order, even the tensor perturbation becomes gauge
dependent.

In particular, the second-order tensor perturbations generated from the
quadratic combinations of linear scalar perturbations (induced tensor
perturbations) must depend on the gauge condition, as the linear scalar
perturbations depend on the choice of the constant-time hypersurface. There
are studies of such induced tensor modes in the literature
\citep{Mollerach-etal-2004,Baumann-etal-2007,Ananda-etal-2007,Arroja-etal-2009,Assadullahi-Wands-2009,Assadullahi-Wands-2010,Jedamzik-etal-2010,Alabidi-etal-2013,Saga/etal:2010},
but {\it all} these studies have been based on one particular gauge condition,
the zero-shear gauge in our terminology. Note that, with no entirely clear
reason, in the literature, this gauge condition is often termed as the
longitudinal, Newtonian, conformal-Newtonian, or Poisson gauge, etc.  The
zero-shear gauge takes the scalar part of the shear of the normal frame vector to vanish; if we ignore the vector and tensor perturbation, this statement is
valid for fully nonlinear orders in perturbation, see Eqs.\ (B7) and (C7) in
\citet{Hwang-Noh-2013}.

In this work, we explicitly show the gauge dependence of the power spectrum of
the induced tensor perturbations. The main results are summarized in
Eq.\ (\ref{h-PS}) in a unified form, and shown in Figure \ref{fig:GW-PS}.
This paper is organized as following. In Section \ref{sec:Equations}, we
introduce our notations and basic equations. In Section \ref{sec:Gauge}, we
use the nonlinear gauge transformation to show the gauge dependence of the
tensor modes to the second order and relations of the tensor-mode solutions
among different gauge conditions, see Eqs.\ (\ref{h_x-GT}) and
(\ref{eq:h-solutions}). In Section \ref{sec:Fourier}, we present the tensor
power spectrum in several gauge conditions in a unified form. We discuss
the implication of our result to the future observations in Section
\ref{sec:Discussion}. We set $c \equiv 1$.

\section{Equations}
                                 \label{sec:Equations}

As the metric convention we have
\bea
   & &
       d s^2 = -  a^2 \left( 1 + 2 A \right) d \eta^2
       - 2 a^2 B_{i} d \eta d x^i
       + a^2 \left( \gamma_{ij} + 2 C_{ij} \right) d x^i d x^j,
\eea
where the spatial indices of $B_i$ and $C_{ij}$ are raised and lowered using
the background metric of comoving coordinate $\gamma_{ij}$; for a spatially
flat background with $K = 0$ we have $\gamma_{ij} = \delta_{ij}$; $i,j,k,
\dots$ are the three-dimensional spatial indices and $a,b,c, \dots$ are spacetime indices. We decompose
the spatial vector $B_i$ and the spatial tensor $C_{ij}$ into the scalar,
vector and tensor perturbations as \citep{York-1973}
\bea
   & &
       A \equiv \alpha, \quad
       B_i \equiv \beta_{,i} + B^{(v)}_i, \quad
       C_{ij} \equiv \varphi \gamma_{ij}
       + \gamma_{,i|j}
       + C^{(v)}_{(i|j)}
       + h_{ij},
   \label{metric-decomposition}
\eea
where $B^{(v)|i}_i \equiv 0 \equiv C^{(v)|i}_i$ (transverse vector)
 and $h^j_{i|j} \equiv 0 \equiv h^i_i$ (transverse-traceless tensor); a
vertical bar indicates a covariant derivative based on $\gamma_{ij}$ as the
metric; like $B_i$ and $C_{ij}$, indices of $B^{(v)}_i$, $C^{(v)}_i$ and
$h_{ij}$ are raised and lowered using $\gamma_{ij}$ as the metric; we define
$A_{(ij)} \equiv {1 \over 2} (A_{ij} + A_{ji})$. We set
$\chi \equiv a (\beta + a \dot \gamma)$; an overdot indicates time derivative
based on cosmic time $t$ (defined with $dt \equiv a d \eta$). For the energy-momentum tensor, we have
\bea
   & &
       \widetilde T_{ab}
       = \widetilde \mu \widetilde u_a \widetilde u_b
       + \widetilde p \left( \widetilde g_{ab}
       + \widetilde u_a \widetilde u_b \right)
       + \widetilde \pi_{ab},
\eea
where $\widetilde \mu$, $\widetilde p$, $\widetilde u_c$ and
$\widetilde \pi_{ab}$ are the energy density, the pressure, the fluid
four-vector and the anisotropic stress, respectively \citep{Ehlers-1993}.
We decompose them as
\bea
   & &
       \widetilde \mu \equiv \mu + \delta \mu, \quad
       \widetilde p \equiv p + \delta p, \quad
       \widetilde u_i \equiv a \Gamma {v_i}, \quad
       v_i \equiv - v_{,i} + v^{(v)}_i, \quad
       \Gamma \equiv - \widetilde n_c \widetilde u^c
       = {1 \over \sqrt{1 - a^2 \overline h^{ij} {v_i v_j}}},
   \nonumber \\
   & &
       \widetilde \pi_{ij} \equiv a^2 \Pi_{ij}, \quad
       \Pi_{ij} \equiv {1 \over a^2} \left(
       \Pi_{,i|j} - {1 \over 3} \gamma_{ij} \Delta \Pi \right)
       + {1 \over a} \Pi^{(v)}_{(i|j)}
       + \Pi^{(t)}_{ij}
       + {1 \over 3} \gamma_{ij} \Pi^k_k.
   \label{v-notation}
\eea
where $\widetilde n_a$ is the normal four-vector and $\Gamma$ is the Lorentz factor. Indices of $v_i$, $v^{(v)}_i$, $\Pi_{ij}$, $\Pi^{(v)}_i$ and $\Pi^{(t)}_{ij}$ are raised and lowered using $\gamma_{ij}$ as the metric. We have $v^{(v)|i}_i \equiv 0 \equiv \Pi^{(v)|i}_i$ and $\Pi^{(t)i}_{\;\;\;\;\;i} \equiv 0 \equiv \Pi^{(t)j}_{\;\;\;\;\; i|j}$; tildes indicate the covariant quantities. We added $\Pi^k_k$ term in the decomposition of $\Pi_{ij}$ as we have $\Pi^k_k \neq 0$ to the nonlinear order.
$\overline h^{ij}$ is the inverse metric of the ADM (Arnowitt-Deser-Misner) metric $\overline h_{ij}$ defined as $\widetilde g_{ij} \equiv \overline h_{ij}$. For $\gamma \equiv 0 \equiv C^{(v)}_i$ and $h_{ij} \equiv 0$ we can derive explicit form of the inverse metric and we have \citep{Hwang-Noh-2013,FNL-Multi}
\bea
   & &
       \overline h_{ij}
       = a^2 \left( 1 + 2 \varphi \right) \gamma_{ij}, \quad
       \overline h^{ij}
       = {1 \over a^2 ( 1 + 2 \varphi )} \gamma^{ij}, \quad
       \Gamma
       = {1 \over \sqrt{1 -{v^k v_k \over 1 + 2 \varphi}}}, \quad
       \Pi^i_i = {v^j v^j \over 1 + 2 \varphi} \Pi_{ij}.
   \label{FNL-1}
\eea
Setting $\gamma \equiv 0 \equiv C^{(v)}_i$ corresponds to taking the spatial gauge condition without losing any generality and without missing any physics: according to Bardeen ``Since the background 3-space is homogeneous and isotropic, the perturbations in all physical quantities must in fact be {\it gauge invariant} under purely spatial gauge transformations.'' \citep{Bardeen-1988}; this statement is true to fully nonlinear order \citep{Noh-Hwang-2004,Hwang-Noh-2013}.

The tracefree ADM propagation equation can be written as
\bea
   & &
       {1 \over a^2} \left( \nabla_i \nabla_j
       - {1 \over 3} \gamma_{ij} \Delta \right)
       \left[ {1 \over a} \left( a \chi \right)^{\displaystyle\cdot}
       - \alpha - \varphi
       - 8 \pi G \Pi \right]
       + {1 \over a} \nabla_{(i} \left\{
       {1 \over a^2} \left[ a^2 \left( B^{(v)}_{j)}
       + a \dot C^{(v)}_{j)} \right) \right]^{\displaystyle\cdot}
       - 8 \pi G \Pi^{(v)}_{j)} \right\}
   \nonumber \\
   & & \qquad
       + \ddot h_{ij}
       + 3 H \dot h_{ij}
       - {\Delta - 2 K \over a^2} h_{ij}
       - 8 \pi G \Pi^{(t)}_{ij}
       = n_{ij},
   \label{eq:ADMpropagation}
\eea
where $n_{ij}$ indicates pure nonlinear parts, see Eq.\ (109) of
\citet{Hwang-Noh-2007}; in our notation we absorb the $\Pi^k_k$-part to the right-hand-side, thus $n_{ij}$ is tracefree.
The tensor part of Eq.~(\ref{eq:ADMpropagation}) becomes
\bea
   & &
       \ddot h_{ij}
       + 3 H \dot h_{ij}
       - {\Delta - 2 K \over a^2} h_{ij}
       - 8 \pi G \Pi^{(t)}_{ij}
       = s_{ij},
   \label{GW-eq}
\eea
with $s_{ij}$ being the transverse-tracefree projection of $n_{ij}$:
\bea
   & &
       s_{ij}
       \equiv {\cal P}_{ij}^{\;\;\; k\ell} n_{k \ell}
       \equiv n_{ij}
		 - {1 \over 3} \gamma_{ij} n^k_k
       + {1 \over 2} \left( \nabla_i \nabla_j
       - {1 \over 3} \gamma_{ij} \Delta \right)
       \left( \Delta + 3 K \right)^{-1}
       \left[ n^k_k - 3 \Delta^{-1}
       \left( n^{k\ell}_{\;\;\;\;|k\ell} \right) \right]
   \nonumber \\
   & & \qquad
       - 2 \nabla_{(i} \left( \Delta + 2 K \right)^{-1}
       \left[ n_{j)|k}^k
       - \nabla_{j)} \Delta^{-1}
       \left( n^{k\ell}_{\;\;\;\;|k\ell} \right) \right],
   \label{s_ij}
\eea
where ${\cal P}_{ij}^{\;\;\; k\ell}$ is a transverse-tracefree projection operator on a symmetric spatial tensor.
For the spatially flat case, $K = 0$ we have
\bea
   & &
       s_{ij}
       = {\cal P}_{ij}^{\;\;\; k\ell} n_{k\ell}
       = n_{ij}
       + {1 \over 2} \left( \nabla_i \nabla_j - \gamma_{ij}
       \right) n^k_k
       - 2 \Delta^{-1} \nabla_{(i} n_{j),k}^k
       + {1 \over 2} \Delta^{-2} \left( \nabla_i \nabla_j + \delta_{ij} \Delta \right)
       n^{k\ell}_{\;\;\;\;,k\ell}.
   \label{s_ij-flat}
\eea
Up to this point the decomposition and equations are valid to fully nonlinear
order in perturbation.

In the following we shall consider only the scalar perturbations as the
source of the tensor modes. As the spatial gauge condition we take the $\gamma \equiv 0$ \citep{Bardeen-1988}. Thus, we have $A = \alpha$, $B_i = \chi_{,i}/a$ and $C_{ij} = \varphi \gamma_{ij}$. Considering the quadratic combination of linear scalar perturbation, Eq.\ (71) of \citet{Hwang-Noh-2007} gives
\bea
   & &
       n_{ij}
       = {1 \over a^3} \left[ a \left( 2 \varphi \chi_{,i|j}
       + \varphi_{,i} \chi_{,j} + \varphi_{,j} \chi_{,i} \right) \right]^{\displaystyle\cdot}
       + {1 \over a^2} \left( \kappa \chi_{,i|j}
       - 4 \varphi \varphi_{,i|j}
       - 3 \varphi_{,i} \varphi_{,j} \right)
       + {1 \over a^4} \left( \chi^{,k}_{\;\;\;|i} \chi_{,j|k}
       - K \chi_{,i} \chi_{,j} \right)
   \nonumber \\
   & & \qquad
       + {1 \over a^2} \left[ 2 \dot \chi_{,i|j} \alpha
       - H \chi_{,i|j} \alpha
       + \chi_{,i|j} \dot \alpha
       - 2 \left( \alpha + \varphi \right) \alpha_{,i|j}
       - \alpha_{,i} \alpha_{,j}
       - 2 \alpha_{,(i} \varphi_{,j)} \right]
       + 8 \pi G \left( \mu + p \right) v_{,i} v_{,j}
   \nonumber \\
   & & \qquad
       - {1 \over 3} \gamma_{ij} \bigg\{
       {1 \over a^3} \left[ a \left( 2 \varphi \Delta \chi
       + 2 \varphi^{,k} \chi_{,k} \right) \right]^{\displaystyle\cdot}
       + {1 \over a^2} \left( \kappa \Delta \chi
       - 4 \varphi \Delta \varphi
       - 3 \varphi^{,k} \varphi_{,k} \right)
       + {1 \over a^4} \left( \chi^{,k|\ell} \chi_{,k|\ell}
       - K \chi^{,k} \chi_{,k} \right)
   \nonumber \\
   & & \qquad
       + {1 \over a^2} \left[ 2 \alpha \Delta \dot \chi
       - H \alpha \Delta \chi
       + \dot \alpha \Delta \chi
       - 2 \left( \alpha + \varphi \right) \Delta \alpha
       - \alpha^{,k} \alpha_{,k}
       - 2 \alpha^{,k} \varphi_{,k} \right]
       + 8 \pi G \left( \mu + p \right) v^{|k} v_{,k}
       \bigg\}.
   \label{n_ij-general}
\eea

For later convenience, here we summarize the basic set of linear order scalar perturbation equations, see Eqs.\ (95)-(101) in \citet{Hwang-Noh-2007}:
\bea
   & & \kappa \equiv 3 H \alpha - 3 \dot \varphi
       - {\Delta \over a^2} \chi,
   \label{eq1} \\
   & & 4 \pi G \delta \mu + H \kappa
       + {\Delta + 3 K \over a^2} \varphi
       = 0,
   \label{eq2} \\
   & & \kappa + {\Delta + 3 K \over a^2} \chi
       - 12 \pi G \left( \mu + p \right) a v
       = 0,
   \label{eq3} \\
   & & \dot \kappa + 2 H \kappa
       - 4 \pi G \left( \delta \mu + 3 \delta p \right)
       + \left( 3 \dot H + {\Delta \over a^2} \right) \alpha
       = 0,
   \label{eq4} \\
   & & \dot \chi + H \chi - \varphi - \alpha
       - 8 \pi G \Pi
       = 0,
   \label{eq5} \\
   & & \delta \dot \mu
       + 3 H \left( \delta \mu + \delta p \right)
       - \left( \mu + p \right) \left( \kappa - 3 H \alpha
       + {\Delta \over a} v \right)
       = 0,
   \label{eq6} \\
   & & {\left[ a^4 \left( \mu + p \right) v \right]^\cdot
       \over a^4 \left( \mu + p \right)}
       - {1 \over a} \alpha
       - {1 \over a \left( \mu + p \right)} \left(
       \delta p + {2 \over 3} {\Delta + 3 K \over a^2} \Pi \right)
       = 0.
   \label{eq7}
\eea
$\kappa$ is a perturbed part of the trace of extrinsic curvature.

\subsection{Zero-shear gauge}

In the zero-shear gauge we set $\beta \equiv 0 \equiv \gamma$, thus
$\chi = 0$. For $\Pi = 0$, we have $\alpha = - \varphi$ to the linear order, and Eq.\ (\ref{n_ij-general}) gives
\bea
   & &
       n_{ij \chi}
       = - {1 \over a^2} \left[ 4 \varphi \varphi_{,i|j}
       + 2 \varphi_{,i} \varphi_{,j}
       - {1 \over 3} \gamma_{ij} \left( 4 \varphi \Delta \varphi
       + 2 \varphi^{|k} \varphi_{,k} \right) \right]
       + 8 \pi G \left( \mu + p \right) \left(
       v_{,i} v_{,j}
       - {1 \over 3} \gamma_{ij} v^{|k} v_{,k} \right),
   \label{n_ij-ZSG-no-Pi}
\eea
where the sub-index $\chi$ indicates the zero-shear gauge. In the zero-shear gauge we have $\alpha = \alpha_\chi$, $\varphi = \varphi_\chi$, and $v = v_\chi$. To the linear order, from Eqs.\ (\ref{eq1}) and (\ref{eq3}), we have
\bea
   & &
       4 \pi G \left( \mu + p \right) a v = - \dot \varphi - H \varphi.
\eea
These equations are valid for general $K$.

In the matter dominated era the growing solutions to the linear order for $K = 0 = \Lambda$ are \citep{MDE-1994}
\bea
   & &
       \varphi_\chi = {3 \over 5} C, \quad
       \alpha_\chi = - {3 \over 5} C, \quad
       \kappa_\chi = - {9 \over 5} H C, \quad
       v_\chi = - {2 \over 5} {1 \over a H} C, \quad
       \delta_\chi = {6 \over 5} \left( 1 - {1 \over 3} {\Delta \over a^2 H^2} \right) C,
   \label{ZSG-lin-sol}
\eea
where $C$ is a constant (in time) coefficient indicating the growing solution.

\subsection{Comoving gauge}

In the comoving gauge we set $v \equiv 0 \equiv \gamma$. In the zero-pressure case with $\Pi = 0$ and $K = 0$, to the linear order, from Eqs.\ (\ref{eq1})-(\ref{eq7}), we have
\bea
   & &
       \alpha = 0, \quad
       \dot \varphi = 0,  \quad
       {1 \over a} \left( a \chi \right)^{\displaystyle\cdot} = \varphi, \quad
       \kappa = - {\Delta \over a^2} \chi.
\eea
Using this, Eq.\ (\ref{n_ij-general}) gives
\bea
   & &
       n_{ij v}
       = {1 \over a^2} \left( \kappa \chi_{,i|j}
       - 2 \varphi \varphi_{,i|j}
       - \varphi_{,i} \varphi_{,j} \right)
       + {1 \over a^4} \chi^{,k}_{\;\;\;|i} \chi_{,j|k}
       - {1 \over 3} \delta_{ij} \left[
       {1 \over a^2} \left( \kappa \Delta \chi
       - 2 \varphi \Delta \varphi
       - \varphi^{,k} \varphi_{,k} \right)
       + {1 \over a^4} \chi^{,k|\ell} \chi_{,k|\ell} \right],
   \label{n_ij-CG}
\eea
where the sub-index $v$ indicates the comoving gauge. In this gauge we have $\alpha = \alpha_v$, $\varphi = \varphi_v$, $\chi = \chi_v$, and $\kappa = \kappa_v$.

In the matter dominated era the growing solutions to the linear order for $K = 0 = \Lambda$ are \citep{MDE-1994}
\bea
   & &
       \varphi_v = C, \quad
       \alpha_v = 0, \quad
       \chi_v = {2 \over 5} {1 \over H} C, \quad
       \kappa_v = - {2 \over 5} {\Delta \over a^2 H} C, \quad
       \delta_v = - {2 \over 5} {\Delta \over a^2 H^2} C.
   \label{CG-lin-sol}
\eea
The normalization of the growing solution is based on the conserved nature of a variable $\varphi_v \equiv C$.

\section{Gauge Issue}
                                 \label{sec:Gauge}

We consider the gauge transformation $\widehat x^a = x^a + \xi^a (x^e)$. To the linear order, using $\xi_i \equiv {1 \over a} \xi_{,i} + \xi^{(v)}_i$ with $\xi^{(v)|i}_i \equiv 0$, we have [see Eq.\ (250) in \citet{Noh-Hwang-2004}]:
\bea
   & &
       \widehat \alpha = \alpha
       - {1 \over a} \left( a \xi^0 \right)^\prime, \quad
       \widehat \beta = \beta - \xi^0
       + \left( {1 \over a} \xi \right)^\prime, \quad
       \widehat B^{(v)}_i = B^{(v)}_i + \xi^{(v)\prime}_i, \quad
       \widehat \gamma = \gamma - {1 \over a} \xi, \quad
       \widehat C^{(v)}_i = C^{(v)}_i - \xi^{(v)}_i,
   \nonumber \\
   & &
       \widehat \varphi = \varphi - {a^\prime \over a} \xi^0, \quad
       \widehat \chi = \chi - a \xi^0, \quad
       \widehat \kappa = \kappa + \left( 3 \dot H + {\Delta \over a^2} \right) a \xi^0, \quad
       \widehat v = v - \xi^0, \quad
       \widehat \delta = \delta - {\mu^\prime \over \mu} \xi^0
       = \delta + 3 ( 1 + w) {a^\prime \over a} \xi^0,
   \label{GT-linear}
\eea
where the prime indicates a time derivative based on $\eta$ and $x^0 = \eta$; $w \equiv p/\mu$.
From these we have constructed gauge-invariant combinations
\bea
   & &
       \chi_v \equiv \chi - a v, \quad
       \chi_\varphi \equiv \chi - {1 \over H} \varphi, \quad
       \chi_\kappa \equiv \chi + {\kappa \over 3 \dot H + {\Delta \over a^2}}, \quad
       \chi_\delta \equiv \chi
       + {\delta \over 3 (1 + w) H}\,,
   \label{GI-linear1}
\eea
which correspond to $\chi$ (the scalar part of the shear of the normal frame vector) in, respectively, comoving gauge ($\chi_v$),
uniform expansion gauge ($\chi_\varphi$),
uniform curvature gauge ($\chi_\kappa$),
and uniform density gauge ($\chi_\delta$), and
\bea
       \varphi_v \equiv \varphi - a H v, \quad
       \varphi_\chi \equiv \varphi - H \chi\,,
   \label{GI-linear2}
\eea
are the scalar metric (curvature) perturbation $\varphi$ in comoving gauge ($\varphi_v$)
and zero-shear gauge ($\varphi_\chi$).

To the linear order we {\it fix} the spatial (including the
scalar and vector) gauge by conditions \citep{Bardeen-1988}
\bea
   & &
       \gamma \equiv 0 \equiv C^{(v)}_i.
   \label{spatial-gauge}
\eea
Under these gauge conditions, the spatial gauge degrees of freedom are fixed
completely with $\xi = 0 = \xi^{(v)}_i$, thus $\xi_i = 0$, to the linear order. The second order gauge transformation is given from Eq.\ (231) in
\citet{Noh-Hwang-2004} as
\bea
   \widehat C_{ij}
   &=&
       C_{ij}
       - {a^\prime \over a} \xi^0 \gamma_{ij}
       - \xi_{(i|j)}
       + B_{(i} \xi^0_{,j)}
       - \left( C_{ij}^\prime + 2 {a^\prime \over a} C_{ij} \right) \xi^0
       - {1 \over 2} \xi^0_{,i} \xi^0_{,j}
       + \xi^0 \left[ {a^\prime \over a} \xi^{0 \prime}
       + {1 \over 2} \left( {a^{\prime\prime} \over a} + {a^{\prime 2} \over a^2} \right) \xi^0 \right]
       \gamma_{ij}
   \nonumber \\
   &\equiv&
       C_{ij}
       - {a^\prime \over a} \xi^0 \gamma_{ij}
       - \xi_{(i|j)}
       + {\cal C}_{ij},
\eea
where ${\cal C}_{ij}$ indicates pure quadratic parts of the gauge transformation property of $C_{ij}$. From this, using the decomposition in Eq.\ (\ref{metric-decomposition}), we can show
\bea
   & &
       \widehat \varphi
       = \varphi - {a^\prime \over a} \xi^0
       + {1 \over 2} \left( \Delta + 3 K \right)^{-1}
       \left[ \left( \Delta + 2 K \right) {\cal C}^k_k
       - {\cal C}^{k\ell}_{\;\;\;|k\ell} \right],
   \label{GT-varphi} \\
   & &
       \widehat \gamma
       = \gamma - {1 \over a} \xi
       - {1 \over 2} \left( \Delta + 3 K \right)^{-1} \left[ {\cal C}^k_k
       - 3 \Delta^{-1} \left( {\cal C}^{k\ell}_{\;\;\;|k\ell} \right) \right],
   \label{GT-gamma} \\
   & &
       \widehat C^{(v)}_i
       = C^{(v)}_i
       - \xi^{(v)}_i
       + 2 \left( \Delta + 2 K \right)^{-1}
       \left[ {\cal C}^k_{i|k}
       - \nabla_i \Delta^{-1} \left( {\cal C}^{k\ell}_{\;\;\;|k\ell} \right) \right],
   \label{GT-C} \\
   & &
       \widehat h_{ij}
       = h_{ij}
       + {\cal P}_{ij}^{\;\;\; k\ell} {\cal C}_{k\ell}.
   \label{GT-h_ij}
\eea
To the second order we can continue taking the spatial and
rotational gauge by the same conditions in Eq.\
(\ref{spatial-gauge}). These are possible by suitable choices of
$\xi$ and $\xi^{(v)}_i$ using Eqs.\ (\ref{GT-gamma}) and
(\ref{GT-C}); i.e., the spatial gauge conditions to the second order determine $\xi$ and $\xi^{(v)}$ to the second order as
\bea
   & &
       \xi
       = - {a \over 2} \left( \Delta + 3 K \right)^{-1}
       \left[ {\cal C}^k_k
       - 3 \Delta^{-1} \left( {\cal C}^{k\ell}_{\;\;\;\;|k\ell} \right)
       \right], \quad
       \xi^{(v)}_i
       = 2 \left( \Delta + 2 K \right)^{-1}
       \left[ {\cal C}^k_{i|k}
       - \nabla_i \Delta^{-1} \left( {\cal C}^{k\ell}_{\;\;\;\;|k\ell} \right)
       \right].
\eea
Notice that even in the case of vanishing vector perturbation we should not ignore $\xi^{(v)}_i$ to the second order. By taking conditions in Eq.\ (\ref{spatial-gauge}), we have
\bea
   & &
       {\cal C}_{ij}
       = \left( {1 \over a} \chi_{,(i} + \Psi^{(v)}_{(i} \right)
       \xi^0_{,j)}
       - {1 \over 2} \xi^0_{,i} \xi^0_{,j}
       - \left( h_{ij}^\prime + 2 {a^\prime \over a} h_{ij} \right) \xi^0
       + \xi^0 \left[ - \varphi^\prime - 2 {a^\prime \over a} \varphi
       + {a^\prime \over a} \xi^{0 \prime}
       + {1 \over 2} \left( {a^{\prime\prime} \over a} + {a^{\prime 2} \over a^2} \right) \xi^0 \right] \gamma_{ij},
\eea where \bea
   & &
       \chi \equiv a \left( \beta + \gamma^\prime \right), \quad
       \Psi^{(v)}_i \equiv B^{(v)}_i + C^{(v) \prime}_i,
\eea are spatially gauge-invariant combinations to the linear order.
We can show that the $\gamma_{ij}$ part in ${\cal C}_{ij}$ does not
affect the tensor mode gauge transformation in Eq.\ (\ref{GT-h_ij}).

Now, we consider pure scalar perturbation to the linear order.
Ignoring the $\gamma_{ij}$ part that do not contribute to the tensor modes,
we have
\bea
   & &
       {\cal C}_{ij}^{(\rm tensor)} = {1 \over a} \chi_{,(i} \xi^0_{,j)}
       - {1 \over 2} \xi^0_{,i} \xi^0_{,j}.
   \label{cal-Cij}
\eea
Using Eqs.\ (\ref{GT-linear})-(\ref{GI-linear2}) and Eq.\ (\ref{cal-Cij}),
we can construct a set of variables ${\cal C}_{ij{\rm x}}$
such that the gauge
transformation is given with ${\cal C}_{ij}^{\rm (tensor)}$,
\bea
   & &
		\widehat {\cal C}_{ij{\rm x}}
       = {\cal C}_{ij{\rm x}}
       - {\cal C}_{ij}^{(\rm tensor)}
\eea
where ${\rm x} = \chi$, $v$, $\varphi$, $\kappa$, and $\delta$.
The explicit expressions for $C_{ij\rm {x}}$ that we consider here are
given as following:
\bea
   & &
       {\cal C}_{ij v}
       \equiv {1 \over a} \chi_{,(i} v_{,j)} - {1 \over 2} v_{,i} v_{,j}, \quad
       {\cal C}_{ij \chi} \equiv {1 \over 2 a^2} \chi_{,i} \chi_{,j}, \quad
       {\cal C}_{ij \delta} \equiv
       - {1 \over a^2} {\chi_{,(i} \delta_{,j)} \over 3 ( 1 + w ) H}
       - {1 \over 2 a^2} {\delta_{,i} \delta_{,j}
       \over [3 ( 1 + w ) H]^2},
   \nonumber \\
   & &
       {\cal C}_{ij \varphi} \equiv {1 \over a^2 H}
       \chi_{,(i} \varphi_{,j)}
       - {1 \over 2 a^2 H^2} \varphi_{,i} \varphi_{,j}, \quad
       {\cal C}_{ij \kappa} \equiv - {1 \over a^2} \chi_{,(i}
       \left( {\kappa \over 3 \dot H + {\Delta \over a^2}} \right)_{,j)}
       - {1 \over 2 a^2}
       \left( {\kappa \over 3 \dot H + {\Delta \over a^2}} \right)_{,(i}
       \left( {\kappa \over 3 \dot H + {\Delta \over a^2}} \right)_{,j)}.
\eea
Note that, unlike the gauge-invariant variables $\chi_{\rm x}$ and
$\varphi_{\rm x}$ that we have defined earlier,
${\cal C}_{ij {\rm x}}$ is not a gauge-invariant notation.
With these new variables, we can show that
\bea
   & &
       {\cal C}_{ij v} - {\cal C}_{ij \chi}
       = - {1 \over 2 a^2} \chi_{v,i} \chi_{v,j}
       = - {1 \over 2} v_{\chi,i} v_{\chi,j}, \quad
       {\cal C}_{ij \varphi} - {\cal C}_{ij \chi}
       = - {1 \over 2 a^2} \chi_{\varphi,i} \chi_{\varphi,j},
   \nonumber \\
   & &
       {\cal C}_{ij \kappa} - {\cal C}_{ij \chi}
       = - {1 \over 2 a^2} \chi_{\kappa,i} \chi_{\kappa,j}, \quad
       {\cal C}_{ij \delta} - {\cal C}_{ij \chi}
       = - {1 \over 2 a^2} \chi_{\delta,i} \chi_{\delta,j},
   \label{C-relation}
\eea
thus
\bea
   & &
       {\cal C}_{ij {\rm x}} - {\cal C}_{ij \chi}
       = - {1 \over 2 a^2} \chi_{{\rm x},i} \chi_{{\rm x},j}.
\eea

Using these notations, we can construct a unified form of explicit gauge-invariant combination $h_{ij {\rm x}}$ as
\bea
   & &
       h_{ij {\rm x}}
       = h_{ij}
       + {\cal P}_{ij}^{\;\;\; k\ell} {\cal C}_{k\ell {\rm x}},
\eea
where $h_{ij {\rm x}}$ is a unified notation of the gauge-invariant combinations; for example, for ${\rm x} = \chi$, $h_{ij\chi}$ is a gauge-invariant combination which is the same as $h_{ij}$ in the zero-shear gauge ($\chi \equiv 0$), and similarly for ${\rm x} = v$ (the comoving gauge), ${\rm x} = \varphi$ (the uniform-curvature gauge), ${\rm x} = \kappa$ (the uniform-expansion gauge), and ${\rm x} = \delta$ (the uniform-density gauge). Therefore, using the relations in Eq.\ (\ref{C-relation}) we arrive at a rather simple relation between the tensor perturbations in a general gauge ${\rm x}$ and the ones in the zero-shear gauge (with subscript $\chi$) as,
\bea
       h_{ij {\rm x}} - h_{ij \chi}
       = - {1 \over 2 a^2}
       {\cal P}_{ij}^{\;\;\; k\ell} \chi_{{\rm x},k} \chi_{{\rm x},\ell}.
   \label{h_x-GT}
\eea
Complete sets of solutions for $\chi_{\rm x}$ in all fundamental gauges are presented in Tables of \citet{MDE-1994} for a pressureless medium and Tables of \citet{IF-1993} for an ideal fluid medium.

\section{Fourier analysis}
                                 \label{sec:Fourier}

We consider a spatially {\it flat} background, where the plane wave solutions
are eigenfunctions of the Laplacian operator. It is, then, convenient to work
in the Fourier space.
We introduce a Fourier decomposition of a tensor perturbation $h_{ij}({\bf x},t)$ as \citep{Mollerach-etal-2004}
\bea
   & &
       h_{ij} ({\bf x}, t)
       = {1 \over (2 \pi)^3} \int d^3 k e^{i {\bf k} \cdot {\bf x}}
       \left[ h ({\bf k}, t) e_{ij} ({\bf k})
       + \overline h ({\bf k}, t) \overline e_{ij} ({\bf k}) \right],
\eea
where
\bea
   & &
       e_{ij} ({\bf k})
       \equiv {1 \over \sqrt{2}} \left[ e_i ({\bf k}) e_j ({\bf k})
       - \overline e_i ({\bf k}) \overline e_j ({\bf k}) \right], \quad
       \overline e_{ij} ({\bf k})
       \equiv {1 \over \sqrt{2}} \left[ \overline e_i ({\bf k}) e_j ({\bf k})
       + e_i ({\bf k}) \overline e_j ({\bf k}) \right],
\eea
are polarization bases for spin-2 fields; we construct the bases from two
transverse unit vectors $e_i$ and $\overline{e}_i$
(satisfying $|e_i| \equiv 1 \equiv |\overline e_i|$ and
$e^i k_i = \overline e^i k_i = e^i \overline e_i = 0$). Note that
the two polarization bases are orthogonal
($e^{ij}({\bf k}) \overline e_{ij}({\bf k}) = 0$) and normalized as
$e^{ij}({\bf k}) e_{ij}({\bf k}) = \overline e^{ij}({\bf k}) \overline e_{ij}({\bf k}) = 1$. These two bases are sometimes called
$e^+_{ij}=\sqrt{2}\,e_{ij}$ and $e^\times_{ij}=\sqrt{2}\,\overline{e}_{ij}$
in literature; see, for example, \citet{dai/etal:2012}.
By using the orthogonality of the polarization bases, we have
\bea
   & &
       h ({\bf k}, t)
       = e^{ij} ({\bf k}) \int d^3 x e^{- i {\bf k} \cdot {\bf x}}
       h_{ij} ({\bf x}, t), \quad
       \overline h ({\bf k}, t)
       = \overline e^{ij} ({\bf k}) \int d^3 x e^{- i {\bf k} \cdot {\bf x}}
       h_{ij} ({\bf x}, t).
\eea
For the spatially flat case ($K=0$) and in the absence of the genuine
tensor-origin contribution to anisotropic stress $\Pi_{ij}^{(t)}$, which is
the case for the standard cosmological models,
the gravitational wave equation in Eq.\ (\ref{GW-eq}) becomes
\bea
   & &
       \left( \partial_t^2 + 3 H \partial_t + {k^2 \over a^2} \right)
       h ({\bf k}, t)
       = e^{ij} ({\bf k}) \int d^3 x e^{- i {\bf k} \cdot {\bf x}}
       s_{ij} ({\bf x}, t)
       = e^{ij} ({\bf k}) \int d^3 x e^{- i {\bf k} \cdot {\bf x}}
       n_{ij} ({\bf x}, t)
       \equiv {1 \over a^2} s ({\bf k}, t)\,,
   \label{h_k-eq}
\eea
with $s_{ij}$ given in terms of $n_{ij}$ in Eq.\ (\ref{s_ij-flat}).
The other polarization mode $\overline h({\bf k},t)$ obeys the equation
parallel to Eq.~(\ref{h_k-eq}) with $\overline s({\bf k},t)$ defined with
$\overline e_{ij}$ instead of $e_{ij}$.
In the parity-preserving Universe, the two polarization modes
$h({\bf k},t)$ and $\overline h({\bf k},t)$ must have the \emph{exactly} same
statistical properties. We, therefore, shall focus only on $h({\bf k},t)$ in
what follows.
The effect of $\overline h({\bf k},t)$ will be taken into account by simply
adding the same contribution at the end of the calculation.

We transform Eq.~(\ref{h_k-eq}) by introducing the new variable
$\mathpzc{v} \equiv ah$ and using the conformal-time derivative (denoted by
prime) as
\bea
   & &
       \mathpzc{v}^{\prime\prime}({\bf k},\eta)
       + \left[ k^2 - {a^{\prime\prime}(\eta) \over a(\eta)} \right]
       \mathpzc{v}({\bf k},\eta)
       = a(\eta) s({\bf k},\eta).
\label{eq:v2nd}
\eea
The solution is then given by
\bea
\mathpzc{v}({\bf k},\eta)
=
\int_{0}^\eta
a(\tilde{\eta}) s({\bf k},\tilde{\eta}) \mathpzc{g}(k;\eta,\tilde\eta)
d\tilde\eta,
\label{eq:v2nd_soln}
\eea
by using the Green's function
\bea
\mathpzc{g}(k;\eta,\tilde\eta)
=
\frac{\mathpzc{v}_1(\eta)\mathpzc{v}_2(\tilde\eta)-\mathpzc{v}_1(\tilde\eta)\mathpzc{v}_2(\eta)}{\mathpzc{v}_1'(\tilde\eta)\mathpzc{v}_2(\tilde\eta)-\mathpzc{v}_1(\tilde\eta)\mathpzc{v}_2'(\tilde\eta)},
\eea
for $\eta\ge\tilde\eta$ and 0 otherwise, as the scalar source at time
$\tilde\eta$, $s({\bf k},\tilde\eta)$, only affects the gravitational waves at
later times $\eta\ge\tilde\eta$. Here, $\mathpzc{v}_1$ and $\mathpzc{v}_2$
are two linearly independent solutions for the homogeneous part of
Eq.~(\ref{eq:v2nd}).

In the flat, matter dominated Universe ($K=0=\Lambda$), we have
$a\propto\eta^2$, and the solutions to the linear order gravitational waves
are \citep{Lifshitz-1946,Weinberg-1972}
\bea
   & &
       \mathpzc{v} = a h \quad \propto \quad x j_1 (x), \quad x y_1 (x) \quad
       = \quad - \cos{x} + {\sin{x} \over x}, \quad
       - \sin{x} - {\cos{x} \over x},
\eea
where $x \equiv k \eta$. The tensor amplitude $\mathpzc{v}$ is gauge-invariant
in the linear order: that is, it is independent of the gauge condition taken
for the scalar perturbation. The induced tensor amplitudes appear from the
second order as Eq.~(\ref{eq:v2nd_soln}), where the Green's function is given as
\bea
   & &
       g_k(\eta, \tilde \eta)
       \equiv {x \tilde x \over k} \left[ j_1 (\tilde x) y_1 (x)
       - j_1 (x) y_1 (\tilde x) \right].
\eea
The induced tensor amplitude depends on the temporal gauge chosen for the
scalar perturbations.

In this section, we shall present explicit expressions for the induced tensor
amplitude in various different temporal gauge conditions. In particular, we
shall calculate the Fourier space expression for the source term
$s_{\rm x}({\bf k},\eta)$ with a temporal gauge condition denoted by the
subscript ${\rm x}$.
We start from the zero-shear gauge and find $s_{\chi}({\bf k},t)$ in the
matter-dominated universe, and generalize to the other gauges by using the
gauge transformation that we have presented in Sec.~\ref{sec:Gauge}.

\subsection{Zero-shear gauge}

In the zero-shear gauge, using Eqs.\ (\ref{n_ij-ZSG-no-Pi}) and (\ref{h_k-eq}) we can show \bea
   & &
       \left( \partial_t^2 + 3 H \partial_t + {k^2 \over a^2} \right)
       h_\chi ({\bf k}, t)
   \nonumber \\
   & & \qquad
       = {1 \over (2 \pi)^3} \int d^3 q
       \left[ e^{ij} ({\bf k}) q_i q_j \right]
       \left[ {2 \over a^2} \varphi_\chi ({\bf q}, t) \varphi_\chi ({\bf k} - {\bf q}, t)
       + 8 \pi G \left( \mu + p \right) v_\chi ({\bf q}, t) v_\chi ({\bf k} - {\bf q}, t)
       \right]
       \equiv {1 \over a^2} s_\chi ({\bf k}, t).
\eea
In the matter dominated era, using Eq.\ (\ref{ZSG-lin-sol}) we have \bea
   & &
       s_\chi ({\bf k})
       = {6 \over 5} {1 \over (2 \pi)^3}
       \int d^3 q \left[ e^{ij} ({\bf k}) q_i q_j \right]
       C ({\bf q}) C ({\bf k} - {\bf q}).
   \label{s_chi}
\eea
The general solution to the second order is
\bea
   & &
       h_\chi (k, \eta) = {s_\chi ({\bf k}) \over k^2}
       + {1 \over a} \left[ c_1 x j_1 (x) + c_2 x y_1 (x) \right].
   \label{h_chi-sol}
\eea
Imposing the initial condition $h_\chi = 0 = h_\chi^\prime$ at $\eta = 0$ we have $c_2 = 0$ and \citep{Mollerach-etal-2004}
\bea
   & &
       h_\chi ({\bf k}, \eta)
       = {s_\chi ({\bf k}) \over k^2}
       \left( 1 + 3 {x \cos{x} - \sin{x} \over x^3} \right)
       \equiv {s_\chi ({\bf k}) \over k^2} g (k \eta).
   \label{h_chi}
\eea
For $x \ll 1$ we have $g = {1 \over 10} x^2$ thus $h_\chi = {1 \over 10} s_\chi \eta^2$. For $x \gg 1$ we have $g = 1$ thus $h_\chi = s_\chi/k^2$.

\subsection{Unified expression in other gauges}

Solutions in other gauge conditions simply follow from the one in the
zero-shear gauge, as Eq.~(\ref{h_x-GT}) gives
\bea
   & &
       h_{\rm x} ({\bf k}, t)
       = h_\chi ({\bf k}, t)
       - {1 \over 2 a^2} {1 \over (2 \pi)^3} \int d^3 q
       \left[ e^{ij} ({\bf k}) q_i q_j \right]
       \chi_{\rm x} ({\bf q}, t) \chi_{\rm x} ({\bf k} - {\bf q}, t).
   \label{h-from-GT}
\eea
The linear solutions in matter dominated era for $K = 0 = \Lambda$ are [see Table 1 of \citet{MDE-1994}]
\bea
   & &
       \chi_v = {2 \over 5} {1 \over H} C, \quad
       \chi_\varphi = - {3 \over 5} {1 \over H} C, \quad
       \chi_\kappa = - {9 \over 5} {H \over 3 \dot H
       + {\Delta \over a^2}} C, \quad
       \chi_\delta = {2 \over 15} {1 \over H}
       \left( 3 - {\Delta \over a^2 H^2} \right) C,
\eea
which are $\chi$ value evaluated in, respectively,
comoving gauge ${\rm x}=v$,
uniform curvature gauge ${\rm x}=\varphi$,
uniform expansion gauge ${\rm x}=\kappa$, and
uniform density gauge ${\rm x}=\delta$.
In the case of the comoving gauge (${\rm x}=v$), we can check that the solution of $h_v$ derived from the gauge transformation in Eq.\ (\ref{h-from-GT}) coincides with solution directly derived from Eqs.\ (\ref{GW-eq}), (\ref{s_ij-flat}) and (\ref{n_ij-CG}).

From Eqs.\ (\ref{s_chi}), (\ref{h_chi}) and (\ref{h-from-GT}) we have the unified expression for the pure second order contribution,
\bea
   & &
       h_{\rm x} ({\bf k}, \eta)
       = {6 \over 5} {1 \over k^2}
       {1 \over (2 \pi)^3}
       \int d^3 q \left[ e^{ij} ({\bf k}) q_i q_j \right]
       C ({\bf q}) C ({\bf k} - {\bf q}) W_{\rm x} ({\bf k}, {\bf q}, \eta),
   \label{eq:h-solutions}
\eea
where
\bea
   & &
       W_\chi = g (k \eta), \quad
       W_v = g (k \eta) - {1 \over 15} {k^2 \over a^2 H^2}, \quad
       W_\varphi = g (k \eta) - {3 \over 20} {k^2 \over a^2 H^2},
   \nonumber \\
   & &
       W_\kappa
       = g (k \eta) - {1 \over 15} {k^2 \over a^2 H^2}
       \left( 1 + {2 \over 9} {q^2 \over a^2 H^2} \right)^{-1}
       \left( 1 + {2 \over 9} {|{\bf k} - {\bf q}|^2 \over a^2 H^2} \right)^{-1},
   \nonumber \\
   & &
       W_\delta
       = g (k \eta) - {1 \over 15} {k^2 \over a^2 H^2}
       \left( 1 + {1 \over 3} {q^2 \over a^2 H^2} \right)
       \left( 1 + {1 \over 3} {|{\bf k} - {\bf q}|^2 \over a^2 H^2} \right).
	\label{eq:Wx}
\eea
We have ${1 \over aH} = {1 \over 2} \eta$. For $x \ll 1$ we have $g = {1 \over 10} x^2$ thus $h_\chi \propto h_{\rm x}$. Thus, we have $h_{\rm x} = 0 = h_{\rm x}^\prime$ at $\eta = 0$. For $x \gg 1$ we have $g = 1$ thus $h_\chi \ll h_{\rm x}$ except for ${\rm x} = \kappa$.

\subsection{Power spectrum: unified expression}

Using the definition of power spectra
\bea
   & &
       \langle C({\bf k}) C({\bf k}^\prime) \rangle
       \equiv (2 \pi)^3 \delta^D ({\bf k} + {\bf k}^\prime ) P_C (k), \quad
       \langle h_{\rm x} ({\bf k}, \eta) h_{\rm x} ({\bf k}^\prime, \eta) \rangle
       \equiv (2 \pi)^3 \delta^D ({\bf k} + {\bf k}^\prime ) \frac12 P_{h_{\rm x}} (k, \eta),
\label{eq:def_Ph}
\eea
we have the expression for the induced tensor power spectrum as
\bea
   & &
       P_{h_{\rm x}} (k, \eta)
       = {144 \over 25} {1 \over k^4} {1 \over (2 \pi)^3}
       \int d^3 q \left[ e^{ij} ({\bf k}) q_i q_j \right]^2
       P_C (q) P_C (|{\bf k} - {\bf q}|)
       W_{\rm x}^2 ({\bf k}, {\bf q}, \eta).
   \label{h-PS}
\eea
Note that the factor $1/2$ in  Eq.~(\ref{eq:def_Ph}) accounts for the two
polarization modes whose power spectrum must be equal.
Here, we assume that primordial curvature perturbations follow Gaussian
statistics. Note that, although it is the same order, the cross-term
multiplying linear order and third order scalar perturbations is not present
because there is no linear order scalar contribution to the induced tensor mode
in the standard Friedmann-Robertson-Walker world models.

\section{Spectrum of induced gravitational waves}
We calculate the spectrum of induced gravitational waves in the standard
$\Lambda$CDM world model adopting the best-fitting cosmological parameters
(maximum likelihood values in the table entitled\\
``{\sf base\_plikHM\_TTTEEE\_lowTEB\_lensing\_post\_BAO\_H080p6\_JLA}'')
from Planck 2015 \citep{planck:2015}:
$\Omega_{\rm b}h^2 = 0.022307$,
$\Omega_{\rm cdm}h^2 = 0.11865$,
$\Omega_{\nu}h^2 = 0.000638$,
$\Omega_\Lambda = 0.69179$,
with current Hubble expansion rate of $H_0=67.78\,{\rm km/s/Mpc}$.
Primordial scalar power spectrum amplitude and spectral index are, respectively,
$\mathcal{A}_s=2.147\times 10^{-9}$ and $n_s=0.9672$ that yield the
normalization of matter power spectrum at present time as $\sigma_8=0.8166$.

From the Einstein equation in the comoving gauge, we find that
\bea
C({\bf k})
= \frac{5}{2}\Omega_{\rm m}\frac{a^2H^2}{k^2}\delta_v({\bf k}),
\eea
which relates the power spectrum of $C$ to the usual linear matter power
spectrum in the comoving gauge $P_L(k)$ as
\bea
P_C(k) = \frac{25}{4}\Omega_{\rm m}^2\left(\frac{a^2H^2}{k^2}\right)^{2}
P_{L}(k).
\label{eq:PCk}
\eea

\begin{figure}
\centering
\includegraphics[width=0.495\textwidth]{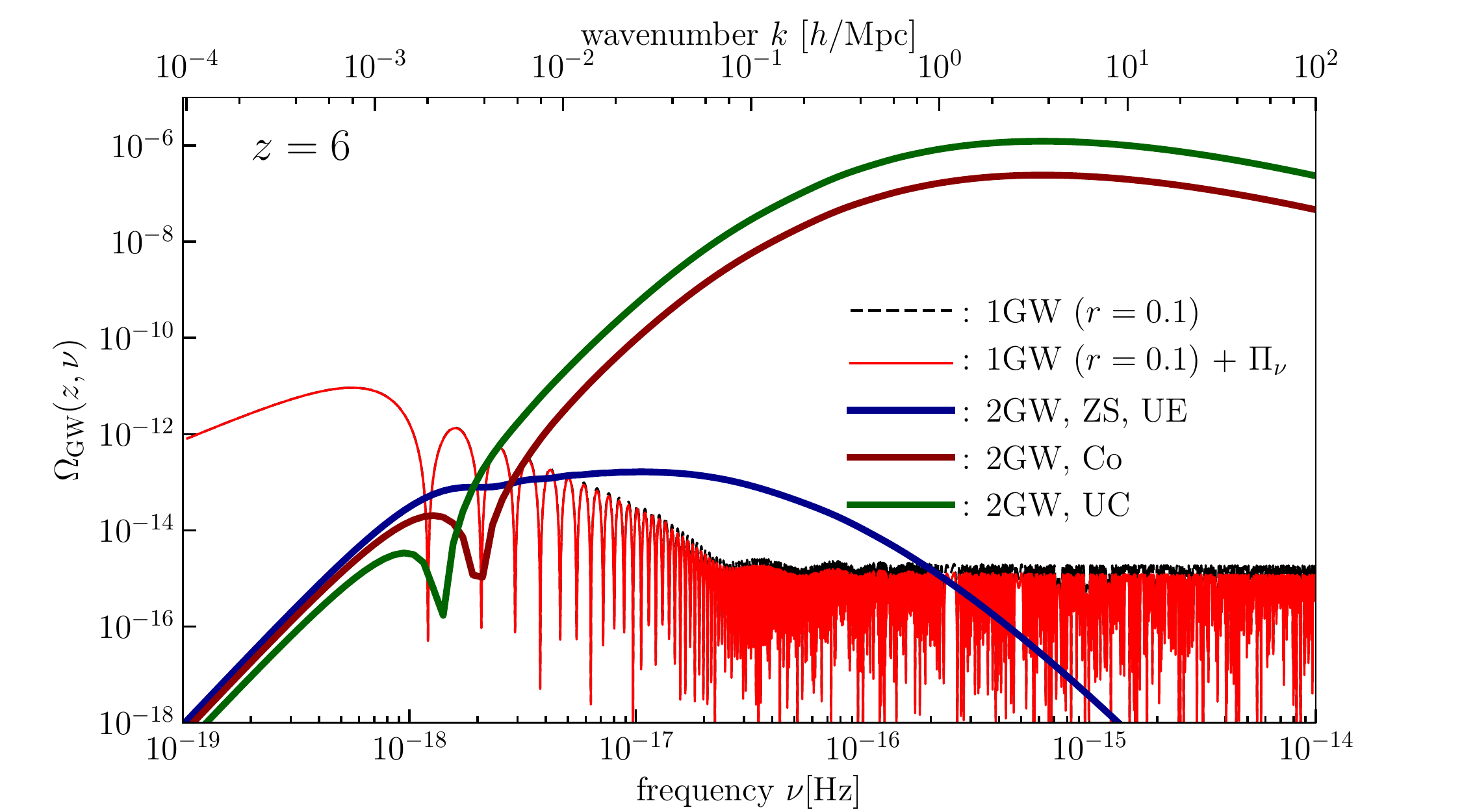}
\includegraphics[width=0.495\textwidth]{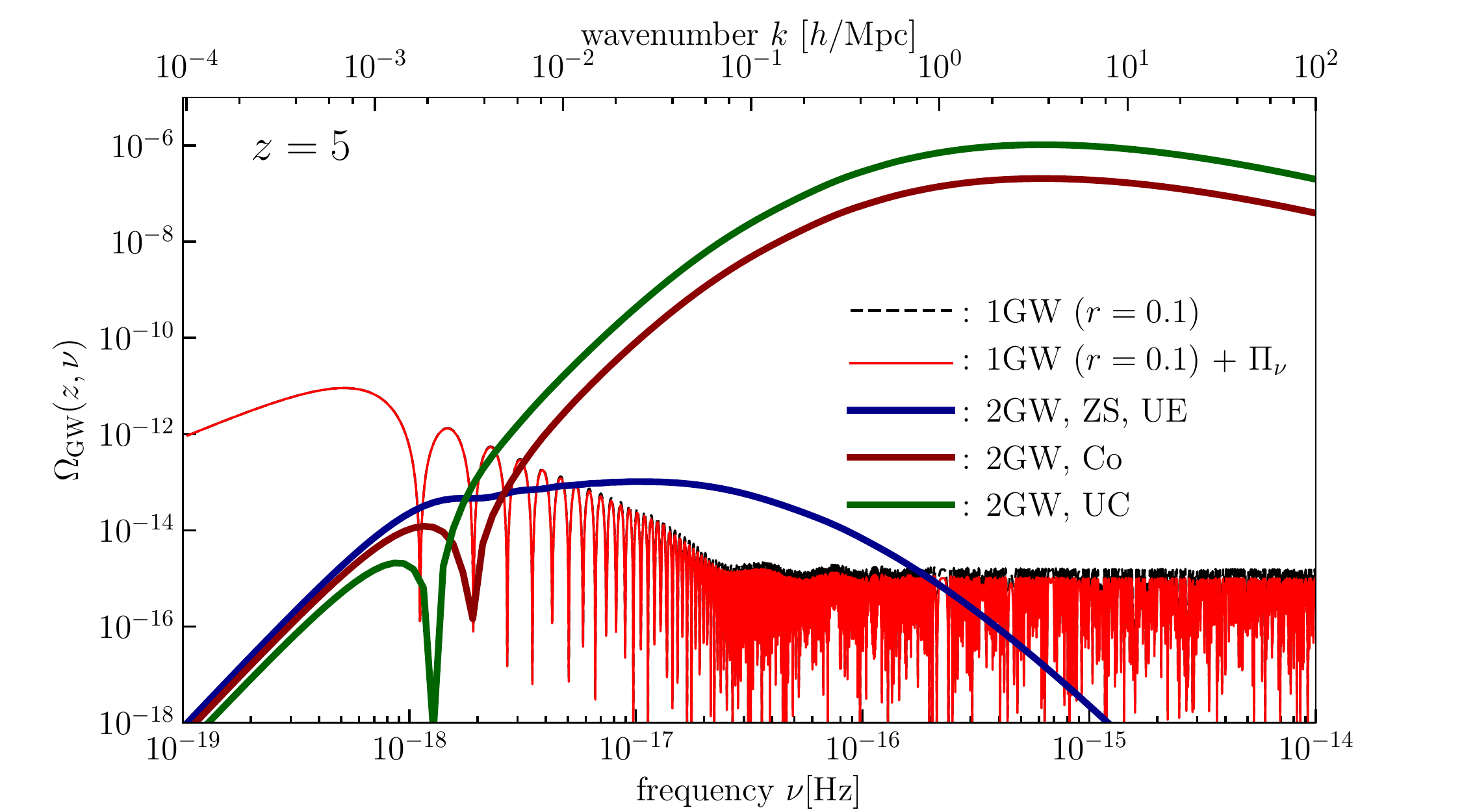}
\includegraphics[width=0.495\textwidth]{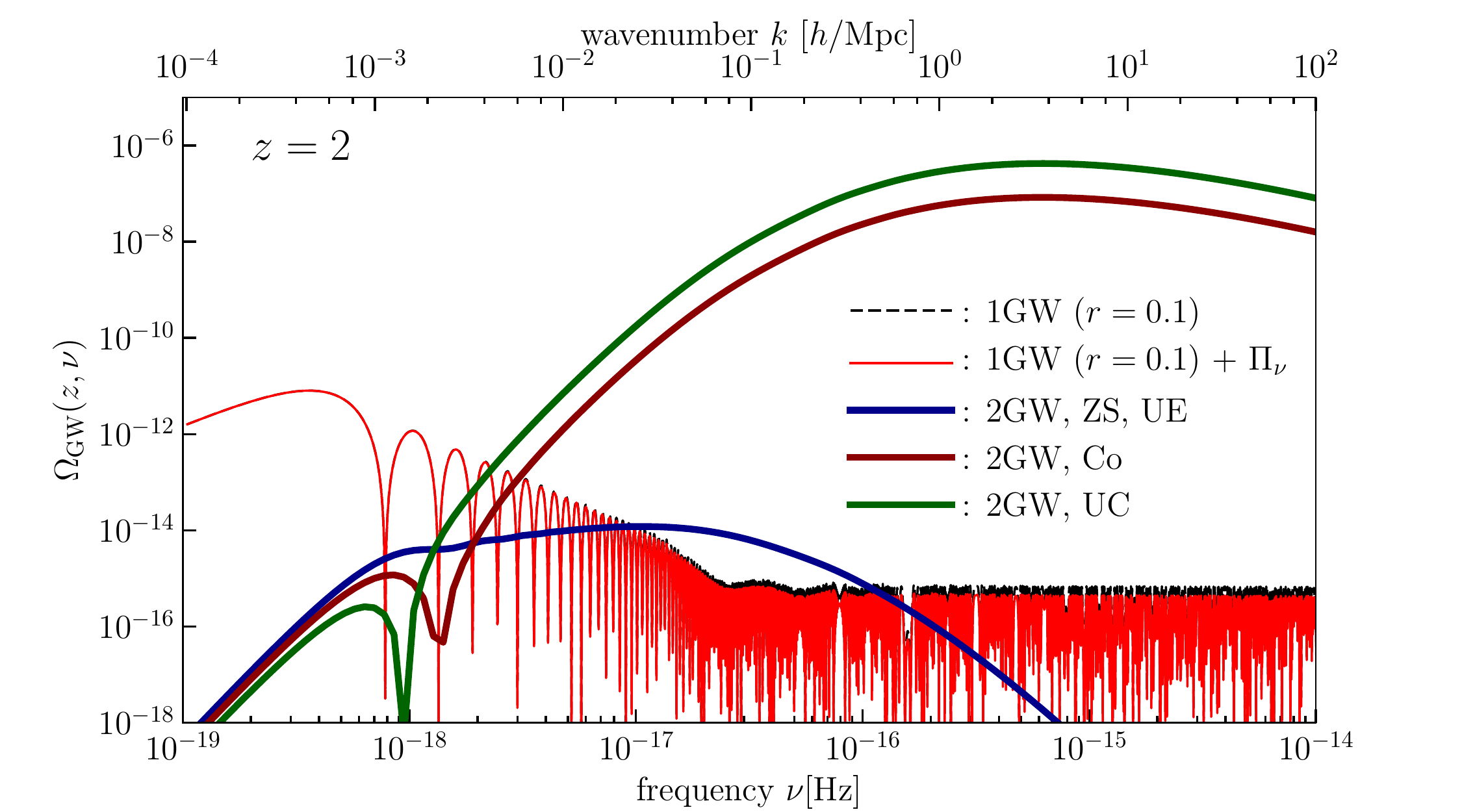}
\includegraphics[width=0.495\textwidth]{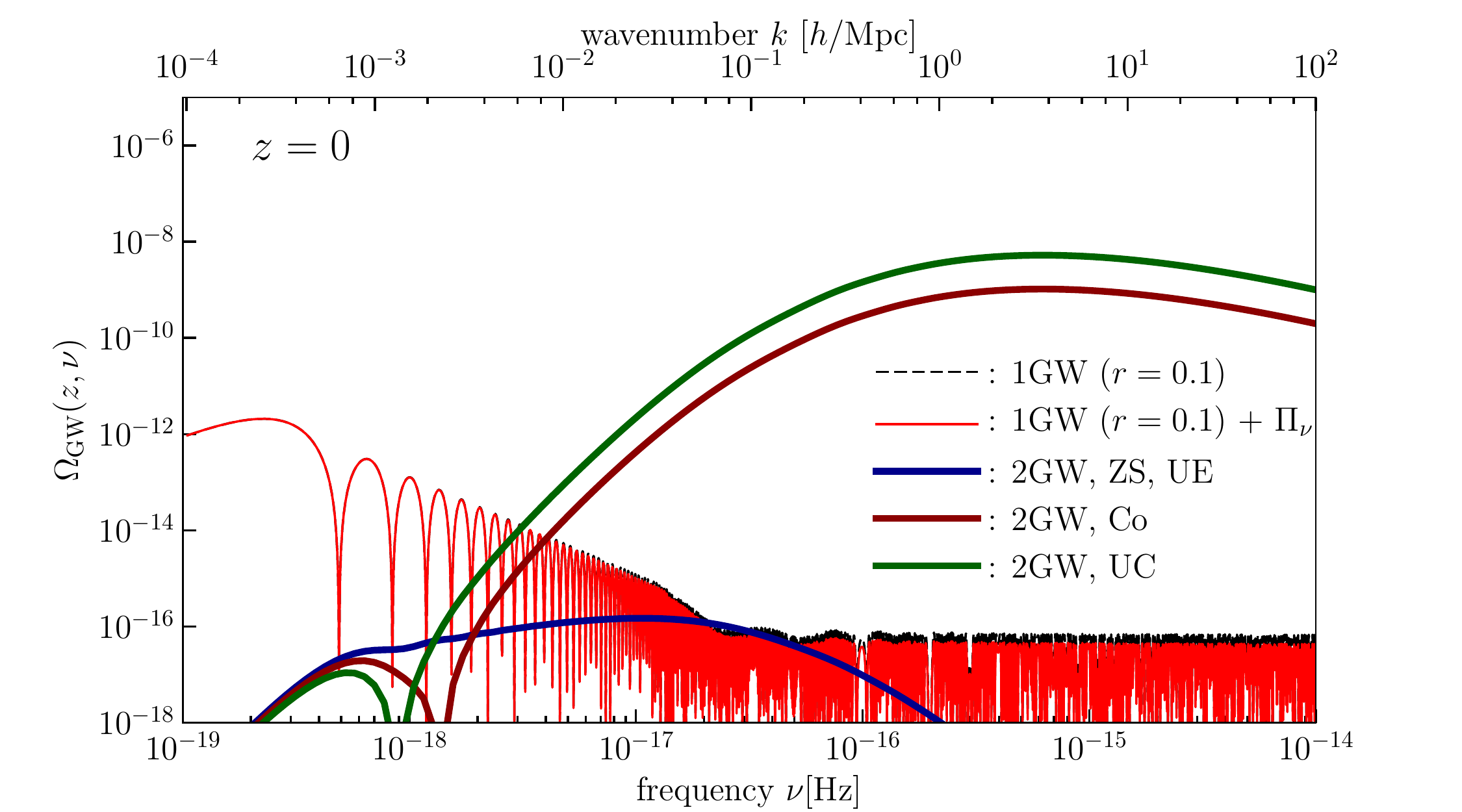}
\caption{
The power spectrum (in the form of GW energy density parameter) for the
 induced tensor-modes (2GW) in four gauge conditions: zero-shear (ZS) gauge and
uniform expansion (UE) gauge ({\it thick, Dark blue}),
comoving (Co) gauge ({\it thick, Dark red}), and
uniform curvature (UC) gauge ({\it thick, dark green}), along with the
linearly evolved primordial gravitational waves (1GW) with ({\it thin, red})
and without ({\it dashed, black}) the damping due to free-streaming neutrinos.
Results from the four redshifts ($z=6$, $z=5$, $z=2$ and $z=0$, from top, left
to bottom, right) are
shown to highlight the time evolution. The induced tensor modes completely
dominate over the primordial signature for $k\gtrsim10^{-2}~[h/{\rm Mpc}]$,
for comoving gauge and uniform curvature gauge, while the induced tensor modes
from zero shear gauge and uniform expansion gauge show moderate excess
between $10^{-2}~[h/{\rm Mpc}]$ and $1~[h/\rm{Mpc}]$.
Searching for the signatures of primordial gravitational waves, therefore,
must take into account the detailed study of the induced tensor modes including
their gauge dependence.
}
\label{fig:GW-PS}
\end{figure}

We have calculated the power spectrum of induced tensor perturbations in
Figure~\ref{fig:GW-PS}, as the gravitational wave energy density parameter
per logarithmic interval
\bea
\Omega_{\rm GW}(k)
\equiv
\frac{1}{12{\cal H}^2}\frac{k^3 P_{h'_{\rm x}}(k)}{2\pi^2}
\simeq
\frac{1}{12{\cal H}^2}\frac{k^5 P_{h_{\rm x}}(k)}{2\pi^2}.
\eea
This follows from the 00-component of the energy-momentum tensor of the
gravitational waves with
$\rho_{\rm GW}\propto \left(h'_{\rm x}\right)^2$ \citep{watanabe/komatsu:2006}.
The second approximated is accurate in sub-horizon scales.
We show both wavenumber ($k$, along the top $x$-axis) and frequency
($f=kc/2\pi$, along the bottom $x$-axis).
Fig.~\ref{fig:GW-PS} shows the induced tensor perturbations calculated in
zero shear (ZS) gauge and uniform expansion (UE) gauge ({\it dark blue}),
comoving gauge (Co, {\it dark red}), and uniform curvature gauge
(UC, {dark green}) at four different redshifts
($z=6$, $z=5$, $z=2$, and $z=0$ from top, left to bottom, right).
We do not present the power
spectrum for the pathological uniform density gauge, because the integrand
in this gauge blows up on small-scales (for larger $k$).
We plot the result down to $k=100~[h/{\rm Mpc}]$, just for the presentation
purpose. Of course, the second-order perturbation theory, breaks down well
before $k\sim 1[h/{\rm Mpc}]$ even for the highest redshift ($z=6$) shown
here; see, for example, \citet{jeong/komatsu:2006}.

First of all, we note the gauge dependence of the power spectra of induced
tensor perturbations. While zero shear gauge and uniform expansion gauge show
the same power spectrum, the power spectra calculated from comoving gauge and
uniform curvature gauge are very different. As the induced tensor modes result
from the non-linear interactions, the power on large-scales is suppressed
and scales as $h_{\rm x}(k)\propto k^2$ in $k\to0$ limit for all cases.
Even on these near-horizon scales, however, amplitude are different for all
cases.

To facilitate the comparison, we have also shown the power spectrum of linear
tensor perturbations with $r=0.1$ with ({\it red solid}) and without
({\it black dotted}) the damping due to free-streaming neutrinos after
neutrino decoupling epoch ($T\simeq 1.5 ~{\rm MeV}$) \citep{weinberg:2004}.
Here, we adopt the damping factor calculation of \citet{watanabe/komatsu:2006}
that the primordial linear gravitational waves $h_{\rm prim}$ are damped
by $80.313\%$ for the modes that re-entering horizon during radiation dominated
epoch. For larger-scale modes, we estimate the damping factor by linearly
re-scaling the small-scale damping factors with the neutrino fraction at the
time of the horizon-crossing. That is, damping factor is applied as
$h^{(1),({\rm with~\nu})}(k,\eta) = [1-0.48582\,\Omega_\nu(\eta=k^{-1})]h^{(1),({\rm without~\nu})}(k,\eta)$, where $h^{(1)}$ is the amplitude of the
linear gravitational waves originated from the primordial Universe.
Note that $\Omega_\nu=0.40523$ during the radiation-dominated epoch, which
gives the desired damping factor of $80.313\%$, but the damping effect dies
out for the modes re-enters the horizon during matter and $\Lambda$-dominated
epoch when the neutrino fraction is negligibly small.

At higher redshift ($z=6$ and $z=5$), the induced tensor power spectrum is
much bigger compared to the linear power spectrum for
$k\gtrsim 10^{-2}~[h/{\rm Mpc}]$, even
for the lowest case for the zero-shear gauge or uniform expansion gauge.
While the amplitude of primordial gravitational waves decays as a linear
theory tensor mode (without the linear source), in the matter dominated epoch,
the amplitude of induced tensor modes stays constant because the induced tensor
mode is proportional to the gravitational potential perturbation which stays
constant.
For lower redshift ($z=2$ and $z=0$), we observe the competition between the
cosmological redshift of the linear tensor mode and the damping of
gravitational potential in the presence of cosmological constant ($\Lambda$).
Note that for the lower redshift, we estimate the induced tensor modes by
simply using the result in the matter-dominated era
[Eqs.~(\ref{eq:h-solutions})-(\ref{eq:Wx})] and re-scale the gravitational
potential power spectrum Eq.~($\ref{eq:PCk}$) with the linear growth factor.

Although the result for zero-shear gauge has been reported in the previous
studies \citep{Mollerach-etal-2004,Baumann-etal-2007,Ananda-etal-2007,Arroja-etal-2009,Assadullahi-Wands-2009,Assadullahi-Wands-2010,Jedamzik-etal-2010,Alabidi-etal-2013,Saga/etal:2010}, the gauge-dependence as well as total
domination of induced tensor modes over primordial gravitational waves
signature is the new result in this work. From these figures, it is clear
that the induced tensor mode contribution must be understood properly
in conjunction with the {\it exact observable} that being considered; that is
the only way to remove the ambiguity due to gauge choice, and, therefore,
to extract truly primordial gravitational waves signature from the
large-scale structure observables.

\section{Discussion}
                                 \label{sec:Discussion}

We have presented the leading order induced tensor power spectrum generated by
the quadratic combination of linear scalar perturbation in the matter
dominated era. The tensor power spectrum depends on the slicing conditions
taken for the linear scalar perturbation. The results are summarized in
Eqs.\ (\ref{eq:h-solutions}) and (\ref{h-PS}) for the solutions and the power
spectra, respectively, in unified forms, and in Figure \ref{fig:GW-PS}.

First of all, we emphasize again that the tensor power spectrum is gauge
dependent as it naturally has to be to the nonlinear order. Comparing the
induced tensor power spectrum with the linearly evolved spectrum of the
primordial gravitational waves (Figure \ref{fig:GW-PS}), even with the
optimistic value of $r=0.1$, we find that the induced tensor power spectrum dominates over the primordial signature for inter-galactic scales
($k\gtrsim10^{-2}~[h/{\rm Mpc}]$). At high redshifts ($z>2$), this is
true for all gauge choices that we have considered in this paper, although
the power spectrum for zero-shear gauge and uniform-expansion gauge decays
faster on smaller scales than comoving gauge and uniform-curvature gauge.

The gauge-dependence that we observe here is a consequence of the
gauge-dependence of the scalar perturbation variables.
In the subhorizon scales, the linear order density perturbation equations
coincide in all four gauges that we consider here \citep{Bardeen-1980,IF-1993}.
The equations for velocity field and gravitational potential, however,
depend on the choice of gauge, and these differences propagate to the
gauge-dependence of the amplitude of the induced gravitational waves.
In the zero-pressure limit the zero-shear gauge and the uniform expansion gauge properly reproduce the {\it exact} (non-perturbative) Newtonian equations for the density, velocity and the gravitational potential \citep{FNL-Newtonian}. If we consider the relativistic pressure, the uniform-expansion gauge is better than the zero-shear gauge \citep{SRHG}. In the sub-horizon scale even to the linear order the uniform-curvature gauge fails to reproduce the Newtonian velocity and gravitational potential \citep{GRG-1999}. The comoving gauge is a curious case. In the zero-pressure limit, the equations for density and velocity exactly coincide with the Newtonian ones even to the second-order perturbations in {\it all} scales but do not have the proper gravitational potential \citep{Noh-Hwang-2004}.

Then, an important question arises: which gauge is the right choice for the observation of induced gravitational waves? As mentioned in Section \ref{sec:Introduction}, in order to
properly address the question one has to specify the observational strategy.
First, the frequency range that we have considered here is
too low (naturally, of order a Hubble time scale,
$f\simeq 10^{-19}-10^{-16}~{\rm Hz}$) to be detected from the direct detection
methods using interferometers such as LIGO or LISA, or pulsar timing array.
Large-scale structure of galaxy distribution offers futuristic, but compelling
methods of detecting tensor modes by, for example, parity-odd (B-mode) part
of the gravitational weak lensing \citep{GWshear}, galaxy clustering
\citep{GWpaper}, cosmic ruler \citep{cosmic_ruler}, as well as clustering
fossils \citep{fossils}.
Because these observables measure the tensor part of the metric perturbations
on scales that we are considering here, and blind about the origin of the
tensor perturbation, we need to understand the induced tensor perturbation
properly in order to pin down the signatures from primordial gravitational
waves. As mentioned earlier, the proper choice of gauge is subject to the exact
way that the tensor perturbations are measured from each observable.

This was true even for the linear order density and velocity power spectra. As
the behavior of density perturbation depends on the gauge
\citep{Lifshitz-1946,Bardeen-1980}, the power spectrum of it should depend on
the gauge as well: in many (but not all) fundamental gauge conditions used in
the literature the behavior of density perturbation happens to coincide far
inside the horizon \citep{Bardeen-1980}. The issue has been resolved in the
density perturbation case by addressing the strategy of measuring the density
power spectrum: by observing the photons traveled from galaxies
\citep{Yoo-etal-2009,Yoo-2010,Bovin-Durrer-2011,Challinor-Lewis-2011,Jeong-etal-2012,Yoo-2014,Jeong-Schmidt-2015}. In the case of cosmic microwave background temperature
anisotropy power spectrum the observational strategy of measuring temperature
`difference' between different angular directions in the sky makes the
observed quantities naturally gauge invariant \citep{Abbott-Wise-1984,Abbott-Schaeffer-1986,Hwang-Noh-1999}.

In this work we only have clarified the gauge dependence of the second order
tensor perturbation power spectrum generated by linear scalar perturbation.
The issue of which one or combination of variable is the right choice for
observed power spectrum is left for future investigation.

\vskip .5cm
%
%
\acknowledgments

J.H.\ was supported by Basic Science Research Program through the National Research Foundation (NRF) of Korea funded by the Ministry of Science, ICT and Future Planning (No. 2016R1A2B4007964).
D.J.\ was supported by National Science Foundation grant AST-1517363.
H.N.\ was supported by National Research Foundation of Korea funded by the Korean Government (No.\ 2015R1A2A2A01002791).

%
%


\end{document}